\documentstyle[twoside,fleqn,epsf,espcrc2]{article}

\newcommand{\AmS}{{\protect\the\textfont2
  A\kern-.1667em\lower.5ex\hbox{M}\kern-.125emS}}

\input epsf.sty
\title{The new definition of lattice gauge fields and the Landau gauge
}
\author{Hideo Nakajima\address{Department of Information Science, Utsunomiya University,\\
2753 Ishii, Utsunomiya 321-8585 Japan (e-mail nakajima@kinu.infor.utsunomiya-u.ac.jp)} and   
        Sadataka Furui\address{School of Science and Engineering, Teikyo University, \\
1-1 Toyosatodai, Utsunomiya 320-8551, Japan (e-mail furui@dream.ics.teikyo-u.ac.jp)
}\thanks{Presenter of the poster.}}

\begin{document}

\begin{abstract}
The Landau gauge fixing algorithm in the new definition of gauge fields
is presented. In this algorithm a new solver of the Poisson equations
based on the Green's function method is used. Its numerical performance of 
the gauge fixing algorithm is presented. Performance of the smeared 
gauge fixing in SU(3) is also investigated.
\end{abstract}

\maketitle

\section{The Landau gauge fixing algorithm
in the new definition of gauge fields
} 
We define the gauge field\cite{DR,NF2}
on links as an element of $SU(3)$ Lie algebra as,
\begin{equation}
e^{A_{x,\mu}}=U_{x,\mu}\ \ ,\ \ \ {\rm where}\ \ A_{x,\mu}^{\dag}=-A_{x,\mu}
\label{DFGAUGE}
\end{equation}
Our definition is different from usual convention\cite{Zw} in $U$-linear form,
\begin{equation}
A^{(1)}={\displaystyle{1\over 2}}(U-U^{\dag})|_{\rm traceless\ part}
\end{equation}
and it is more natural as an averaged gauge field on the link in the continuum
theory. We use the analytic method of
eigenspace projection operator together with Cardano's eigenvalue
solver for transformation between $U$ and $A$.
It is remarkable that the definition (\ref{DFGAUGE}) of gauge field
allows the same type of formulation of the lattice Landau or Coulomb
gauge condition as the continuum theory, i.e., minimizing condition
of $\|A\|^2$ along the gauge orbits\cite{MN,SF}. The definion
$A=\log U$ requires a proper choice of branch for traceless $A$,
and $A$ becomes
a nonanalytic function of $U$, and yet $\|A\|^2$ still remains a continuous
function of $U$. This fact justifies the existence of the global minimum
of $\|A^g\|^2$ along the gauge orbit (the minimal Landau gauge condition).

The gauge transformation is
\begin{equation}
e^{A^g_{x,\mu}} = g_{x}^{\dag}e^{A_{x,\mu}} g_{x+\mu}
\label{GTRNS1}
\end{equation}
where $g_x=e^{\epsilon_x}$ and $\epsilon_x$ is a traceless antihermitian 
matrix.

When $\epsilon$ is infinitesimal 
\begin{eqnarray}
\delta A_{x,\mu}&=&({\cal D}\epsilon)_{x,\mu}=f(adj_{A_{x,\mu}})\partial
\epsilon_{x,\mu}+[A_{x,\mu},\bar\epsilon_{x+\mu/2}]\nonumber\\
{\rm where }&&f(x)=\displaystyle{{x/2\over {\rm tanh}(x/2)}}\ ,\ \ 
\partial\epsilon_{x,\mu}=\epsilon_{x+\mu}-\epsilon_{x},\nonumber\\
 {\rm and}&&\bar\epsilon_{x+\mu/2}={1\over 2}
(\epsilon_{x+\mu}+\epsilon_{x})
\label{INFGT}
\end{eqnarray}

Gribov region $\Omega$\cite{Gv,Zw} is defined as a
set of local minimum points on the gauge orbit
F\\
\begin{equation}
\Delta\|A^g\|^2=-2\langle \partial A|\epsilon\rangle+
\langle \epsilon|-\partial {\cal D}|\epsilon\rangle+\cdots
\label{NORM6}
\end{equation}
\begin{equation}
\Omega=\{A|-\partial {\cal D}\ge 0\ ,\ \partial A=0\}\ \ .
\label{NORM8}
\end{equation}

The minimal Landau gauge condition\cite{Gv,Zw} defines the fundamental
modular region by the abosolute minimum along the gauge orbit,
\begin{equation}
\Lambda=\{A|\|A\|^2={\rm Min}_g\|A^g\|^2\}, \qquad
\Lambda\subset \Omega\ \ .
\label{NORM9}
\end{equation}

The equation to solve for the Landau gauge fixing is highly nonlinear,
\begin{equation}
\partial A^g =0\ \ {\rm where}\ g=e^\epsilon\ .
\label{GFEQ1}
\end{equation}
Linearizing this equation, we can solve it iteratively by the Newton
method with an over-relaxation factor $\eta$,
\begin{equation}
\epsilon =(-\partial {\cal D}(A))^{-1}\eta\ \partial A\ \ .
\label{GFEQ2}
\end{equation}

This is a non-Abelian extension of the Fourier acceleration type\cite{DBKK,CM}.
We make a perturbation expansion with respect to $A$ for the inverse of the 
covariant Laplacian, as
\begin{equation}
(-\partial {\cal D}(A))^{-1}\cong\sum_{n=0}^{N_{end}}((-\partial^2)^{-1}\partial
\tilde D)^n(-\partial^2)^{-1}
\label{GFEQ3}
\end{equation}
where $\tilde {\cal D}={\cal D}(A) - \partial$.

Performance check for various $N_{end}$ is given below.

\section{The new solver of the Poisson equation} 

The inversion of the Laplacian is an important problem in numerical analyses,
and performances of various methods have been investigated\cite{CM}.
Efficiency of the methods depends on types of the problem posed, i.e.,
balance of accuracy and CPU time. We report a new trial, use of the Green
function. While high accuracy being maintained, direct use of the Green
function in a single machine computaion requires a work of order $V^2$, 
and is obviously inefficient in larger size $V$. We devise a method that
makes a partial use of the Green function for coarse lattice with
a preconditioning for the Poisson equation.

Let us consider d dimesional
Poisson equation,
\begin{equation}
-\partial^2 \phi=2d(I-A)\phi=\rho ,
\label{POISS1}
\end{equation}
where $A$ stands for an averaging operation on neighboring sites.
A preconditioned equaiton is given as
\begin{equation}
2d(I-A^2)\phi=(I+A)\rho ,
\label{POISS2}
\end{equation}
where $(I-A^2)$ becomes block-diagonal on even(odd)-site.
After solving the even-site equation (\ref{POISS2}), we use (\ref{POISS1})
for $\phi$ on odd-site. 

The Green function for even(odd)-site equation
(\ref{POISS1}) is defined as
\begin{equation}
2d(I-A^2)G=\delta-\displaystyle{{\frac{1}{V/2}}}\ \ .
\label{GRN1}
\end{equation}
we found that a good approximant $\tilde G$ of the Green function $G$ can
be given by the Green function $G_c$ on coarser lattice with volume $V/2^d$, as
\begin{equation}
\tilde G=(a_0+a_1 A^2 + a_2 A^4)\delta -c + b A^6 G_c, 
\label{GRN2}
\end{equation}
where $a_0$, $a_1$, $a_2$, $b$ and $c$ are optimized parameters.
We can add an extra function $\Delta G$ for a higher accuracy correction.
For an accuracy, $10^{-4}$, of the Green function $G$ on $8^3\times16$
lattice, numbers of sites for supports of $G_c$ and $\Delta G$ are
256 and 162, respectively, out of 4096 of full even-site.
Owing to the high accuracy of the above Green function algorithm,
this Poisson equation solver allows the over-relaxation factor $\eta$
to be a larger value, 1.5 to 1.8 in  gauge fixing algorithm 
(\ref{GFEQ2}) and (\ref{GFEQ3}) and gives a 30 \% faster performance
than that in the usual conjugate gradient solver.

The following Fig. 1 shows processs of gauge fixing in use of the Green
function method and the conjugate gradient method.
\begin{figure}[hbt]
\begin{center}
\leavevmode
\epsfysize=150pt\epsfbox{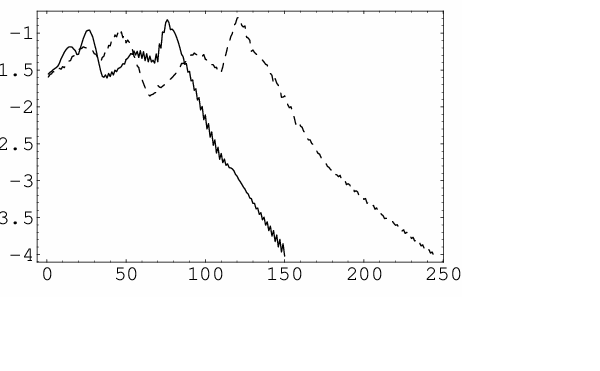}
\end{center}
\caption{A comparison of $log_{10}{\rm Max}\vert \partial A\vert$ in the conjugate gradient
method(dashed) and the Green's function method(continuous).}
\end{figure}

\section{Numerical performance of the gauge fixing algorithm}

We compare performance of the gauge fixing algorithm (\ref{GFEQ2})
and (\ref{GFEQ3}) for various $A$ perturbation orders $N_{end}$ in
Fig. 2, which shows that the non-Abelian extension improves efficiency.
Relative CPU time ratios over that of $N_{end}=3$ case are 1.05 for
$N_{end}=4$ case and 1.91 for $N_{end}=2$ case.
\begin{figure}
\begin{center}
\leavevmode
\epsfysize=150pt\epsfbox{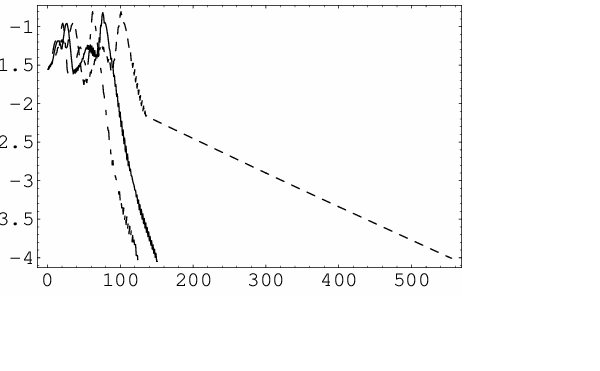}
\end{center}
\caption{The dependence of $log_{10}{\rm Max}\vert \partial A\vert$ on the inclusion of the
 2nd(dashed), 3rd(continuous) and 4th(dash-dotted) order perturbative commutator term 
in the Green's function method.}
\end{figure}

In general, the Fourier accelaration type of methods accomplish much
better performance in ultimate precision over simple
norm-minimizing methods, and our
method is free of critical slowing down.

\section{Smeared gauge fixing in SU(3)}

As is well known, the Landau gauge fixing suffers from the problem of
Gribov's copies, that is, the configuration $A^g$ is captured by the
plenty of local minima of $\|A^g\|^2$ along the gauge orbits.
In order to avoid noise of those Gribov's copies when measuring,
e.g., the gluon propagators etc., the method of gauge
fixing to the global minimum of the $\|A^g\|^2$ 
is necessary. Hetrick and de Forcrand proposed a clever method of
smeared gauge fixing\cite{HdF}. They investigated performance of the
smeared gauge fixing in case of SU(2), and reported that it works well
for large $\beta$. Here we report that it works very well also for SU(3)
in case of large $\beta$. We observed that it works perfectly
in gauge fixing to the global minimum of $\|A^g\|^2$, by using
50 configurations generated by random gauge transformation from a sample
in $\beta=5$ on $4^3\times 8$ lattice, all of which are transformed to the
unique minimum.
\begin{figure}
\begin{center}
\leavevmode
\epsfysize=150pt\epsfbox{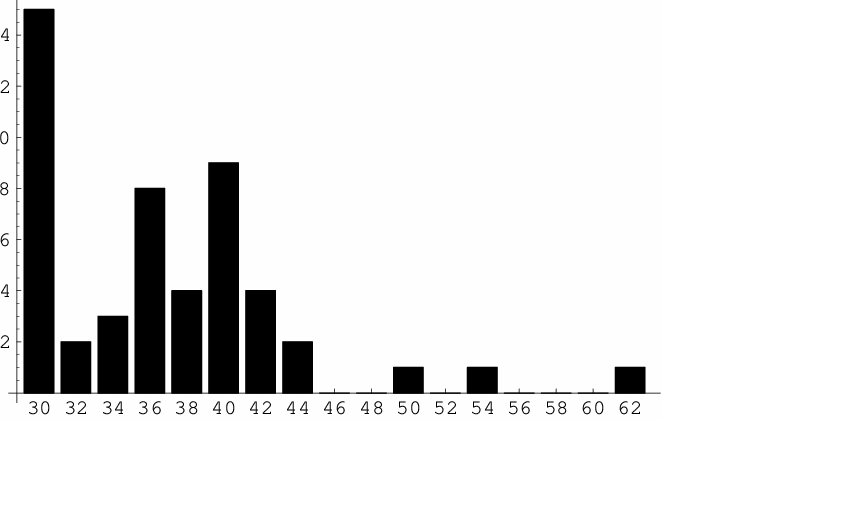}
\end{center}
\caption{A histogram of the number of Gribov copies obtained after the random
gauge transformation. 
Transformation of the 50 samples to the unique minimum $\vert\vert A\vert\vert^2=0.2430$ was achieved.}
\end{figure}

\end{document}